\documentclass[acmsmall]{acmart}

\begin{document}

\title{Expressive MIDI-format Piano Performance Generation}

\author{Jingwei Liu}
\email{jil121@ucsd.edu}
\affiliation{%
  \institution{University of California San Diego}
  \streetaddress{9500 Gilman Drive}
  \city{La Jolla}
  \state{CA}
  \country{USA}
  \postcode{92092}
}

\begin{abstract}
  This work presents a generative neural network that's able to generate expressive piano performance in MIDI format. The musical expressivity is reflected by vivid micro-timing, rich polyphonic texture, varied dynamics, and the sustain pedal effects. This model is innovative from many aspects of data processing to neural network design. We claim that this symbolic music generation model overcame the common critics of symbolic music and is able to generate expressive music flows as good as, if not better than generations with raw audio. One drawback is that, due to the limited time for submission, the model is not fine-tuned and sufficiently trained, thus the generation may sound incoherent and random at certain points. Despite that, this model shows its powerful generative ability in generating expressive piano pieces.
\end{abstract}

\maketitle

\section{Audio versus Symbolic Music Generation}

There is a long history of symbolic music analysis and generation, largely based on traditional music scores and well-notated Western classical music. Nowadays the music scene is gradually shifted towards Afrodiasporic sounds, multicultural hybrid productions, and electronic music which are either facilitated by the multifarious digital tools in DAWs, or live performances based on (collective) improvisations. In the meantime, the restrictions of quantized notes, fixed meters, and orthodox music theories are facing criticisms from modern music practitioners. As a result, to achieve more flexibility and fit in the time, more music-generative systems are developed as audio-based instead of symbolic. Despite all the adversities, the author still insisted on developing this symbolic music generative model based on the following reasons:

\begin{enumerate}
    \item There is still a lack of a good neural network model for music audio generation. Although some sample-based or audio-feature-based neural networks made breakthroughs, we still suffer from the large number of samples in audio format and difficulties in determining higher-level representations that characterize music ontology.
    \item The full potential of symbolic music generation hasn't been achieved yet. Symbolic music is frequently criticized for its stiffness and non-flexibility in generating listening-based music (when playing the symbolic generation from the computer). However, MIDI file type is able to record almost all musical aspects of a performance and reproduce it through the computer. We contend that better data processing is required to minimize information loss, retain most of the musical nature, and generate expressive symbolic music.
    \item As the audio samples are too mechanic and atomic in representing music, finding good high-level representations from audio becomes a necessity. However, the symbolic MIDI data is an optimized high-level musical representation given the audio input it encodes. Therefore, by generating from MIDI, we bypass the feature learning phase and operate the generative system from a more succinct and optimal standpoint.
    \item Music is about taste. There are always arguments on whether an auditory event is music. For most symbolic music, especially classical music, this type of argument is significantly reduced due to its mature music form, long-standing reputation, and the complexity and human intelligence behind it that we now consider masterpieces. However, for many audio streams, it's difficult to assess their quality as music and such data will possibly jeopardize the quality of generation.
\end{enumerate}

\section{Listening-based Data Processing}

\begin{figure}[h]
  \centering
  \includegraphics[width=\linewidth]{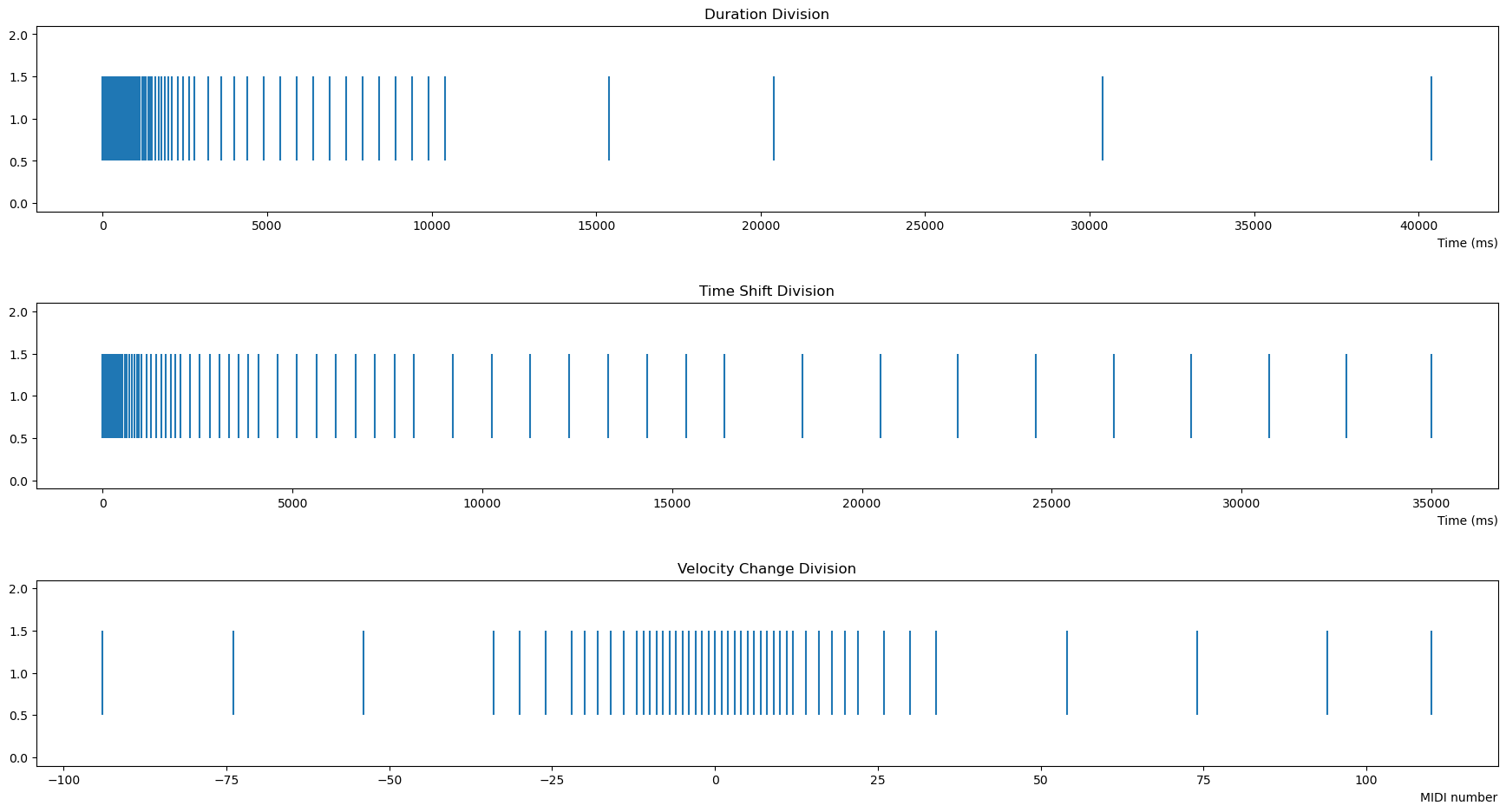}
  \caption{The categorical distributions for given input features. The divisions obey Weber's law where the perceptual changes are proportional to the current values.}
  \label{1}
\end{figure}

Although our generative system is built on symbolic data, the goal is still to generate auditory events from the computer directly, rather than generating notes and scores that need to be interpreted by human performers \cite{performance-rnn-2017}. To achieve this goal, we modified the music data processing in the following ways:
\begin{enumerate}
    \item Abandonment of fixed grid. Many generative systems for Western classical music use metric-based time annotations such as quarter notes. This quantized time notation simplifies the music score but it also loses the expressive micro-timings in the actual performance. Replay sound from quantized time grids makes the piece sound mechanical and nonliving. Here we use note durations measured by milliseconds, which is no longer restricted by the metrical stuff.
    \item A refreshed perspective of monophony and polyphony. Unlike language models, where the words only appear sequentially, music notes can be arranged both horizontally and vertically. The generation of polyphonic music has always been a problem at the center of symbolic music generation. The situation gets more complicated when there is an undetermined number of simultaneous notes in polyphonic music (which is often the case in multi-instrument orchestration or single-instrument piano composition). To tackle this problem, we choose to use another lens to view monophony and polyphony. We claim that there is no real simultaneity of notes. For any two notes that are played by a human performer, there is always a time discrepancy between them, no matter how unnoticeable it is. It means that, since there are no simultaneous events, we can always place the notes sequentially, by their time onsets. This perspective bridges the gap in the perception of monophony and polyphony, which means all polyphonic music can be viewed as monophony in the order of sound attacks. In this sense, we can generate polyphonic music using sequential models. In data processing, we use time shifts to characterize the differences between two consecutive notes.
    \item Not only the notes matter. In almost all symbolic music generative models, the only thing that's being generated and taken into consideration is note. People always use note-related features such as note value, duration, velocity, and onset as inputs and outputs. Reflected on MIDI information, it shows that people always take the note events while ignoring the control events. However, in piano performances, there is one control parameter that plays a role as important as the notes, which is the sustain pedal. In many piano compositions, the sustain pedal is frequently used and makes huge differences if the piece is played without it. The sustain pedal also provides valuable grouping information for the notes. In this work, we take active consideration of the sustain pedal in our generative model, thus the generated music also contains the pedal information.
    \item Mel quantization of auditory features. Except for pedal status, there are 4 note-event inputs, which are note value, duration, velocity, and time shift. The note value and velocity are quantized in MIDI format, and duration and time shift are measured in milliseconds. Instead of equal division, like the Mel spectrogram, we divide the ranges into uneven chunks to better reflect the perceptual truth. We refer to Weber's law for just noticeable differences as our theoretical foundation for the divisions, where the noticeable difference is proportional to the current value. The velocity is featured by its change relative to the previous note instead of its absolute value. The categorical intervals are illustrated in Figure \ref{1}. This uneven quantization not only fits the perceptual reality but also equalizes the data distribution, where a statistic of the entire dataset shows the data is more evenly distributed into these perceptual-based categories.
\end{enumerate}

\section{Convolved Multi-argument LSTM}
\begin{figure}[h]
  \centering
  \includegraphics[width=\linewidth]{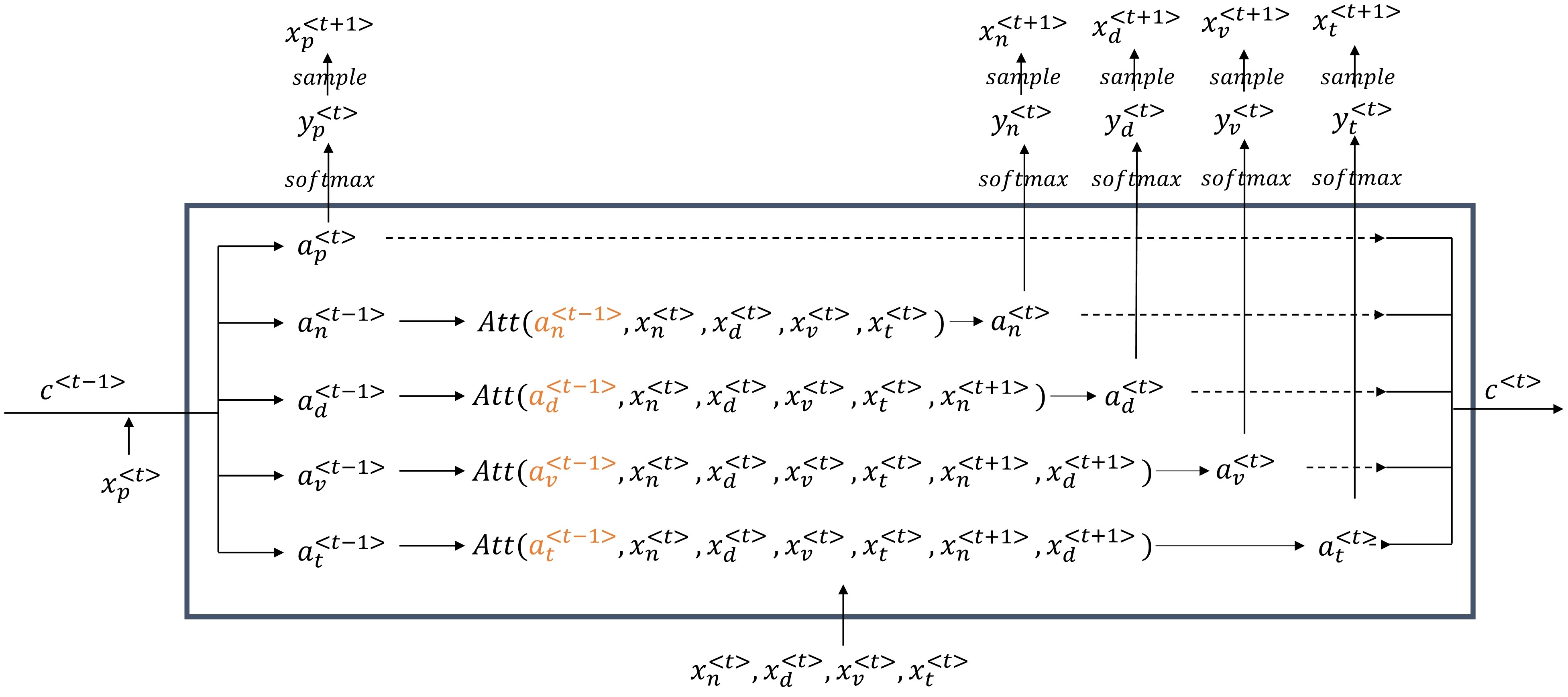}
  \caption{LSTM-Attention cell. A recurrent neural network designed for Multi-input-output generative system.}
  \label{2}
\end{figure}

We use LSTM, the long-short time memory neural network, to model and generate the music sequence, which is supposed to be able to capture long-term dependencies. One problem here is about the multiple inputs and outputs. We have 5 arguments as input features and same for the outputs (which is taken recurrently as next inputs). One consideration is that, these outputs are intercorrelated, which means we cannot generate them as independent heads. To be more specific, if we sample note value and its duration separately, it means no matter what the note value is, the corresponding duration is independent of it, which counters musical intuition. In other words, these outputs should be determined recurrently by taking the generated terms into computation. The author believes there are multiple ways to do it, but she chose to apply the attention mechanism as sub-modules to generate each output. The recurrent cell structure is described in Figure \ref{2}.

\section{Reflections and Comments}
The neural network is built three days before the due date (Nov 12th) so the model is not sufficiently trained (only two epochs). There are lots to fine-tune and adjust in the model to improve its performance so the generation could sound more coherent and "accurate". The preliminary results (the roughly trained parameters) are submitted, which give a head-up on the generative power of the model in generating expressive piano pieces.

This work can be viewed as the author's ultimate endeavor towards the symbolic music generation system. It's considered ultimate because of the limitations of symbolic music in the following ways:
\begin{enumerate}
    \item Non-categorizable instruments. As physical sound sources are more and more difficult to identify in the era of computer-based music production, the categorization of instruments become infeasible. Even if we impose a grand category tag on all instrumental variations from a physical guitar and encode them as guitar, it's no longer possible to recover the music from symbolic information faithfully. When the information loss gets larger than the hearing resolution, the encoding method is no longer adoptable.
    \item Undefinable control events. In addition to the note events, MIDI also records the controls the performer imposed on the instruments. As the controls get multifarious, it's impossible to classify them into 128 categories. This also jeopardizes the recovery ability of the MIDI encodings.
    \item Inconsistent styles and use of instruments. To generate symbolic music data, we need to specify how many instruments we are using and what are they. It's only feasible when we are dealing with very consistent datasets, such as piano performance, choral music, or orchestration with specified and unchangeable instruments. However, these are rare cases. For most music nowadays, let alone DAW productions, even music composed in notational software has undetermined numbers and types of instruments that depend on the composer and nothing restricts them from choosing any instrument. This freedom of instrument genres leaves the symbolic generative model in the dilemma of having to actively determine instruments before generating any notes. This will add great difficulty to the generative models thus making it ill-functional.
\end{enumerate}
\bibliographystyle{ACM-Reference-Format}
\bibliography{sample-base}
\end{document}